\documentclass[12pt]{article}

\newcommand{\be}{\begin{equation}}
\newcommand{\ee}{\end{equation}}
\newcommand{\ba}{\begin{eqnarray}}
\newcommand{\ea}{\end{eqnarray}}

\newcommand{\tr}{\rm tr\; }
 \newcommand{\G}{\Gamma}
 \newcommand{\si}{\sigma}
 \newcommand{\vf}{\varphi}

\newcommand{\Log}{{\rm Log}}

\usepackage{epsf}
\usepackage{epsfig}

\begin{document}
\hoffset=-.4truein\voffset=-0.5truein
\setlength{\textheight}{8.5 in}
\begin{titlepage}
\hfill{LPTENS 09-}

\vskip 0.6 in
{\large   Critical interface: twisting spin glasses at $T_c$}
\vskip .6 in
\begin{center}
 {\bf E. Br\'ezin$^{a)}$} ,  {\bf S. Franz$^{b)}$} ,  {\bf{G.Parisi $^{c)}$}} 
\end{center}
\vskip 5mm
\begin{center}
{$^{a)}$ Laboratoire de Physique
Th\'eorique, Ecole Normale Sup\'erieure}\\ {24 rue Lhomond 75231, Paris
Cedex
05, France. e-mail: brezin@lpt.ens.fr{\footnote{\it
Unit\'e Mixte de Recherche 8549 du Centre National de la
Recherche Scientifique et de l'\'Ecole Normale Sup\'erieure.
}\\
{$^{b)}$ 
Laboratoire de Physique Th\'eorique et Mod\`eles Statistiques\\ 
Universit\'e Paris-Sud 11,
Centre scientifique d'Orsay,
15 rue Georges Cl\'emenceau
91405 Orsay cedex, France
}\\
{$^{c)}$ 
Dipartimento di Fisica, ``Sapienza'' Universit\`a di Roma, P.le A. Moro 2, 00185 Roma, Italy 
2 \\

INFM-CNR SMC, INFN,  ``Sapienza'' Universit\`a di Roma, P.le A. Moro 2, 00185 Roma, Italy }\\
 }}

.
\end{center}
\vskip 3mm {\bf Abstract}  We consider two identical copies of a finite dimensional spin glass coupled at their boundaries. This allows to identify
the analog for a spin glass of twisted boundary conditions in
ferromagnetic system and  it leads to a definition of an interface
free-energy that should scale with a positive power of the system size in
the spin glass phase. In this note we study within mean
field theory the behavior of this
interface at the spin glass critical temperature $T_c$ . 
We show that the leading scaling of the interface free-energy may
be obtained by simple scaling arguments using a cubic field theory of
critical spin glasses and neglecting the replica symmetry breaking
dependence.
\end{titlepage}
\vskip 3mm
 
\section{Introduction}
Sensitivity to boundary conditions is a fundamental tool to study the
nature of the Gibbs states of extended physical systems. If ergodicity
is broken and Gibbs states are not unique, different boundary
conditions can select pure phases or induce interfaces in the
system. For example, in Ising-like ferromagnets in the ferromagnetic
phase, the choice of homogeneous "up'' (respectively "down'') boundary
conditions, where all the spins outside a large region are fixed to
point in the up (resp. down) direction, is enough to select the pure
state with positive (resp. negative) magnetization. Twisted boundary
conditions, where along a specific direction $+$ and $-$ conditions
are chosen at the opposite boundaries (while neutral boundary
condition, e.g. periodic ones, are chosen in the other directions)
induce an interface with positive tension : the free energy cost to impose twisted boundary conditions remains finite in the thermodynamic limit.  On the contrary, in the
paramagnetic phase the system is insensitive to boundaries and the
surface tension is equal to zero.  Using boundary conditions one can
therefore investigate the stability of the low temperature phases and the
lower critical dimension below which the ferromagnetic phase is
unstable.  Around the critical temperature, the behavior of the
interface tension is described by finite size scaling, which implies
that right at the critical temperature, the interface tension
vanishes.
The behavior of the interface free-energy at the 
critical temperature is well understood  in critical fluctuation theory. 
 Below dimension 4 the interfacial free-energy scales with the system size as $\sigma \sim L^{-(D-1)}$, whereas  it scales as $\sigma \sim L^{-3}$ above dimension 4 \cite{Widom}.

 In trying to extend this kind of considerations to study the
 stability of spin glass phases, one is confronted with the fact that pure
 phases are in this case glassy random states, strongly correlated
 with the quenched disorder. Therefore they cannot be selected through
 boundary conditions uncorrelated with the quenched disorder. The
 projection unto some pure phase boundary conditions should depend
 self-consistently on the quenched disorder.  
A possible way to consider disorder-correlated boundary conditions at 
low temperature has been introduced in \cite{BDsg}, where, in reference 
to the Parisi ansatz, two different gauges of the ultrametric matrices were considered at the boundary.

A different procedure, is 
 to consider two "clones'' of a spin glass \cite{FPV} with identical disorder
 coupled at the boundaries. In ref. \cite{FPVint} a situation was
 considered where different values of the overlap (a measure of the
 correlations between configurations) were imposed on the boundaries
 along one direction. There the stability of a low temperature phase,
 with broken replica symmetry (RSB), was investigated against spatial
 fluctuations of the overlap that would restore replica symmetry ;
 this led to a value of the lower critical dimension, below which the
 fluctuations destroy RSB, equal to $D_{LCD}=5/2$. 
In ref. \cite{CGGPV} it was showed that it may be interesting to consider the extreme case where the overlap is 1 at one boundary and -1 (or 1) at the other boundary. These extreme choice simplifies the analysis and it can be implemented in numerical simulations in  a very simple way.
 
 In the present note we consider such a situation at the critical
 temperature itself, where the techniques developed in
 \cite{FPVint} do not apply.  We then compare two types of boundary
 conditions.  We consider two identical clones of cylindrical spin
 glasses (i.e. we impose p.b.c. in the transverse directions) and we
 couple them on the boundaries.  In a first type of b.c.
the configurations of the 
clones are chosen to be identical on both sides of the cylinder. In a second
type of b.c., the configurations of the two systems are still fixed to be identical on one
of the boundaries, but they are opposite in the other boundary.
We study the problem above the upper critical dimension (which is $D_u=6$ for spin glasses \cite{UCD}) where we can neglect loop contributions to the interface free-energy. We find that the interface free-energy scales as $\sigma \sim L^{-5}$ above dimension 6
and it is  natural to conjecture that it scales as $\sigma \sim L^{-(D-1)}$ below that dimension, in analogy with pure systems where it follows from the universality of the free energy  in a correlation volume plus finite size scaling. 


The plan of the paper is the following: In section II we discuss in detail the Ising ordered case. This is well understood and it will guide us in the analysis of the more complex spin glass case. Section III, which constitutes the core of this paper, is devoted to the analysis of the Edwards-Anderson model at $T_c$. Finally we draw some conclusions.

\section{Pure system}
We consider an Ising-like system in a cylindrical geometry. The D-dimensional  cylinder has a length $L$ and a cross-section area $A\sim L^{D-1}$.  The boundary conditions are periodic in the  directions transverse to the axis
of the cylinder, but on the  surfaces  at the end of the longitudinal direction  the spins may be all up or  all down.
Below $T_c$ there is an interfacial tension $\si$ defined as the difference of free energy per unit area
\be \si  =\frac{ F_{\uparrow, \downarrow}- F_{\uparrow, \uparrow}}{A}. \ee
It is a function of the temperature $t\sim(T-T_c)$ , and of the aspect ratio $L/A^{1/(D-1)}$. This tension vanishes near $T_c$ as
\be \si \sim (-t)^{\mu} .\ee
A scaling law due to Widom \cite{Widom} relates $\mu$ to the correlation length exponent $\nu$
\be \mu = \nu (D-1) \ee 
and it may be derived in the standard renormalization group framework \cite{Feng} (for $D\leq 4$). 

Assume that we want to know the behavior of this tension for finite $L$ at $T_c$. We use finite size scaling
\be \si(t,L) = (-t)^{\mu} f(L/\xi) \ee
and, since $\si$ is non singular at $T_c$ for $L$ finite, it implies that
\be f(x) \sim_{x\to 0} x^{-\mu/\nu} .\ee
Therefore at $T_c$ the interfacial tension vanishes with $1/L$ as 
\be \si(0,L) \sim L^{-\mu/\nu} = L^{-(D-1)} .  \ee
In other words at $T_c$ 
\be\label{1}  F_{\uparrow, \downarrow}- F_{\uparrow, \uparrow} \sim L^0. \ee

In appendix A we give the details of the verification of this  behavior (\ref{1}) in a simple Landau theory (i.e. tree level of a $\varphi^4$ theory) :  i.e. for a free energy at $T_c$ 
\be F= A\int_0^L dz ( \frac {1}{2} (\frac{ d\vf}{dz})^2 + \frac{g}{4} \vf^4) \ee
(we have assumed translation invariance in the $(D-1)$ transverse directions). 
Up-up b.c. are defined be the fact that at both boundaries the order parameter is fixed to the value $m$, up-down conditions correspond to impose the value $m$ on the left boundary and $-m$ on the right one. One finds that independently of the values of $m$, the surface tension behaves as  
\be\label{2} \frac{ F_{\uparrow, \downarrow}- F_{\uparrow, \uparrow}}{A} = \frac {C}{gL^3} \ee
with a positive constant $C$ given by 
\be C= \frac{5}{192\pi^2} \G^8(\frac{1}{4}) - \frac{1}{12\pi^{3/2}} \G^6(\frac{1}{4})\simeq 44.8. \ee
Clearly this mean field theory should only be compared with (\ref{1}) in four dimensions, the result (\ref{2}) is indeed in agreement since $A\sim L^3$. 
\section{Spin glass}
In a spin glass twisting fixed boundary conditions, uncorrelated with the disordered
couplings, does not produce any interface since it can
be gauged away in a change of the random couplings. For a given realization of the random disorder It may increase or decrease the free energy.

In this paper, 
 we consider two identical copies of a cylindrical sample of a spin glass with identical couplings $J_{ij}$ and we study  two sets of  boundary conditions as follows :
\begin{itemize}
\item In the up-up situation  the overlap between the spins of the two copies in the left ($x=0$) plane and in the $x=L$ plane are equal to a positive value $l_0$. 

\item In the up-down situation the overlap is still equal $l_0$ in the left plane, but it is  $-l_0$ in the right plane.
\end{itemize}

In ref. \cite{CGGPV} Contucci et al. have recently considered boundary conditions of
this kind in the extreme case $l_0=1$.  Notice that in this case, the construction
leads to effective interactions within the spins in the planes $x=0$
and $x=L$ which are doubled. Choosing the interactions in those planes with
the same statistics as for the other planes, would lead to  effective
interactions between the spins in these planes stronger than the other
interactions. In order to avoid such a strong difference, 
the strength of the interaction among spins on the boundary planes was reduced by half in \cite{CGGPV}.

Our aim here is to characterize the behavior of the difference
$\si=\frac{ F_{\uparrow, \downarrow}- F_{\uparrow, \uparrow}}{A}$ at $T_c$
as a function of $L$. Again we shall limit ourselves to mean field
theory. Our purpose is  to test whether  Widom's scaling law holds in this glassy system, by verifying that above the upper critical 
dimension $D_{UCD}=6$ the interface tension scales as $\sigma\sim L^{-5}$. 

We would like to study this problem within a Landau-like 
 free-energy close to $T_c$ expanded in powers of an order parameter. Unfortunately,
this can not be done directly if the value $l_0$ of the order
parameter at the boundary is chosen to be large. In particular, we
could not directly consider the boundary conditions chosen in \cite{CGGPV}.
However, the small order parameter expansion can be saved even in this
case, arguing that, analogously to the ferromagnetic case, the result
for $\si $ should not depend on the value $|l_0|$ imposed at the
boundaries. We will see that this is true within the Landau model. We
have also confirmed this in a spherical spin-glass model (see Appendix B)
which has a Landau expansion identical to the one of the standard
Ising spin glass \cite{Theo}, and for which one can write equations which are  valid even 
for large values of the order parameter.

We start from a short range standard Edwards-Anderson model \cite{EA}. 
After tracing out over the spins we can rewrite it, in the long wavelength limit,  in terms of a $2n\times 2n$ matrix $Q_{a,b}$ (for the bulk problem we would need only a $n\times n$ matrix $Q$) ;  when $a$ and $b$ are both smaller than $n$, $Q_{ab} $ refers to the overlap between the spins in the first copy ; when the two indexes are larger than $n$ we are dealing with the overlaps $Q_{\alpha, \beta}$ in the second copy and for one index lower than $n$ and one larger , $Q_{a\alpha}$ is a cross-overlap. The critical mean field free-energy  is here

\be \label{free}  F = \int d^Dx  [ \frac{1}{2} \tr (\nabla Q)^2 + \frac{w}{3} \tr Q^3 + \frac{u}{4} \sum_{ab} Q_{ab}^4] \ee
Assuming again translation invariance along the transverse directions $Q (\vec x)  = Q(z)$, we  assume that at $T_c$ the replica-symmetry is unbroken. Therefore we assume
\be Q_{ab} = Q_{\alpha \beta} = q(z) \ee
and 
\be {\label {ansatz}} Q_{a\alpha} = k(z)(1-\delta_{a , \alpha}) + l(z) \delta_{a , \alpha} .\ee
Let us note that the free energy (\ref{free}) is invariant under $S_n\times S_n$ , i.e. independent permutations of the $n$ replicas of system one, and of the $n$ replicas of system two. However  the last term in (\ref{ansatz}) breaks this symmetry down to $S_n$. Indeed one may  imagine the boundary conditions as follows : one adds a coupling $J s^1 s^2$ between the spins of the two systems located in the plane $z=L$. The boundary conditions $s^1 = s^2$ would correspond to $J$ going to $+\infty$, whereas $s^1 = -s^2$ is enforced by the limit $J\to -\infty$. After replicating the  coupling becomes $J\sum s^1_a s^2_a$ and thus the boundary conditions lead to a $\delta_{a, \alpha}$ term in (\ref{ansatz}) \cite{Mezard}

Then one has
\be{ \tr (\frac{dQ}{dz})^2} = 2n(n-1) q'^2 + 2n(n-1)k'^2 +2nl'^2 \ee
\be {\tr Q^3} = 2n(n-1)(n-2) q^3 +6n(n-1)(n-2) qk^2 + 12n(n-1) qkl \ee
\be \sum Q_{ab}^4 = n(n-1) q^4 +  n(n-1) k^4 + n l^4 \ee

and thus, in the zero replica limit, the free energy reads
\be \frac{F}{n} = A\int_0^L dz [ -q'^2 - k'^2 + l'^2 + \frac{w}{3}( 4q^3 +12 qk^2 - 12qkl) + \frac{u}{4} (-q^4-  k^4 +l^4)]  \ee
The quartic term may be dropped at $T_c$ since the functions $q(z), k(z), l(z)$ remain small for large L.  The equations of motion are thus simply

\ba - 2q" +4w(q^2+k^2 -kl) &&= 0 \nonumber \\
 - 2k" + 4w(2qk -ql)&& =0 \nonumber \\
- 2l" +4wqk &&=0 
\label{ode}
\ea

The up-up, and up-down boundary conditions are different in the two cases for the functions $l(z)$ and $k(z)$. In the up-up case, if one imposes the same boundary conditions for $q$ and $k$ , $q(z) = k(z) $ is a solution and we are left with only two equations. There is one constant of motion, namely $ 4q'^2 - l'^2 + \frac{w}{3} (8q^3 -3lq)$, but the last  integration has to be done numerically. In the up-down, one has to deal with three different functions. 

\begin{figure}[ht]
\begin{center}
\includegraphics[width= 0.7 \textwidth]{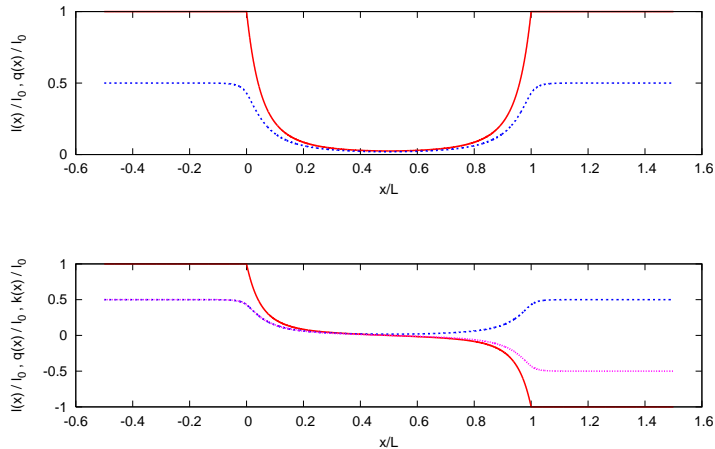}
\caption{
The order parameters in the two solutions 
for the critical 
Edwards-Anderson
 model. Upper panel: the functions $l(x)/l_o$ (red solid curve) and $q(x)/l_0=k(x)/l_0$ (blue dashed curve) as a function of $x/L$ for $v=1000$ and $r=1/2$ 
for up-up boundary conditions.  Lower Panel: 
the functions $l(x)/l_o$ (red solid curve), $q(x)/l_0$ (blue dashed curve) and $k(x)/l_0$ (violet dotted curve)
for up-down boundary conditions and the same values of the parameters.  }
\label{fig1}
\end{center}
\end{figure}

We impose the boundary conditions in the following way: we consider a
sample of size $(1+2 r) L$ with $-rL \le x\le (1+r) L$ and fix the
value of $l(x)$ to preassigned values on the regions $-r L <x<0$, that
we call left boundary, and $L<x<(1+r) L$, that we call right
boundary. In the case of up-up, the value $l_0>0$ is imposed both
at the right and left boundaries, in the up-down case the value $l_0>0$ is
chosen at the left boundary and $-l_0$ is chosen at the right
boundary. The value of $q(x)$ and $k(x)$ on the boundaries are not
fixed by the constraints and they are simply determined by extremization of the
free-energy.  Since $r>0$ and $L$ is large, $q(0)$ and $k(0)$ and
$q(L)$ and $k(L)$ should take the same values that that they would take if a uniform
overlap profile $l(x)=\pm l_0$ for all $x$ was chosen, namely $q(0)=k(0)=q(L)=k(L)=l_0/2$ for up-up conditions and $q(0)=k(0)=q(L)=l_0/2$, $k(L)=-l_0/2$ for up-down conditions.  
In fig. \ref{fig1} is depicted the typical behavior of the various functions in space. 

We expect that, as verified explicitly in the case of the ferromagnet, the interface free-energy does not depend on the imposed value $l_0$. In that case,  
a scaling argument based on the fact that the dominant interaction terms in the free-energy are the cubic ones (namely the transformation $x\to x L$ $q\to w^{-1} L^{-2} q$ and analogously for $k$ an $l$), suggests that the interface tension, in absence of accidental cancellations should behave as $A L^{-5}$ for large $L$, i.e. for $A\sim L^{D-1}$, $\si \sim L^{D-6}$. 
Indeed, one can see by the above rescaling, we can write $F_{\uparrow, \uparrow}=\frac{A}{w^2 L^5}g_{\uparrow, \uparrow}(l_0  w L^2)$,  $F_{\uparrow, \downarrow}=\frac{A}{w^2 L^5}g_{\uparrow, \downarrow}(l_0  w L^2)$ where $g_{\uparrow, \uparrow}$ and $g_{\uparrow, \downarrow}$ are appropriate scaling functions. This is of course true unless some accidental cancellation occurs. If this is the case,  we are indeed authorized to neglect the quartic term. Since this is the term responsible for RSB, we can neglect RSB effects to this leading order. 

\begin{figure}[ht]
\begin{center}
\includegraphics[width= 0.7 \textwidth]{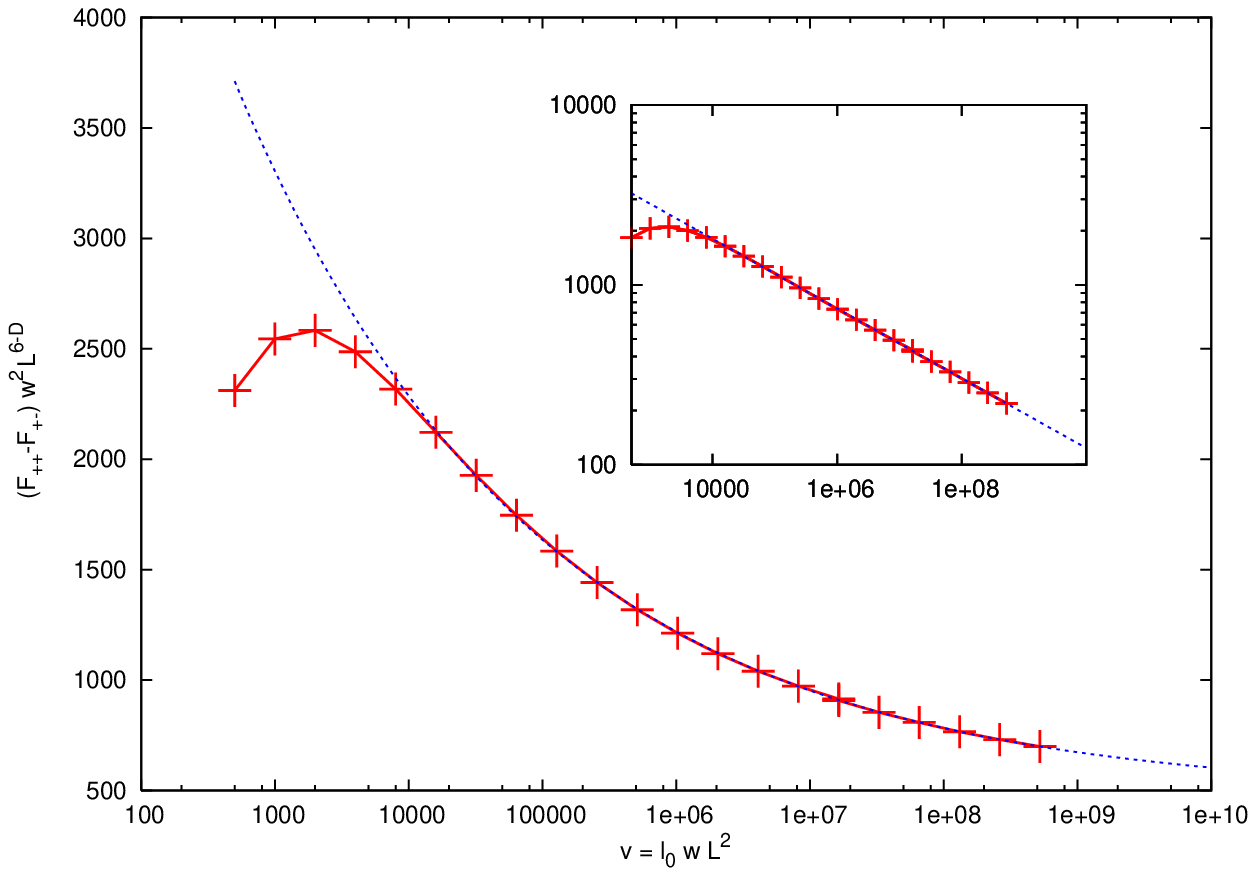}
\caption{The rescaled free-energy difference, $f=(F_{\uparrow, \downarrow}-F_{\uparrow, \uparrow})w^2 L^{6-D}$ as a function of $v=l_o w L^2$, together with a power law fit of the form $f(v)=a+\frac b{v^c}$. 
Inset: log-log plot of $f(v)-a$(red crosses)  together with the same power law fit (blue dotted line) as in the main panel.  Chi-square fitting for $v>10^4$ gives the parameters $a=480\pm 20$, $b=
10790\pm 90$ and $c=0.19\pm 0.01$. }
\label{twist}
\end{center}
\end{figure}

In order to verify that $g_{\uparrow, \downarrow}-g_{\uparrow, \uparrow}$ remains finite for large $L$, we have integrated numerically the rescaled equations (\ref{ode}),  for various values of the parameter $v=l_0 w L^2$. This is done by simple a relaxation method, where the initial profiles are 
iterated until one observes the convergence to a solution of a discretized version of  the equations (\ref{ode}). 
In fig. (\ref{twist}) we have plotted the function $f(v)=g_{\uparrow, \downarrow}(v)-g_{\uparrow, \uparrow}(v)$, together with a power law fit of the kind 
$f(v)=a+\frac b{v^c}$.  While $f(v)$ continues to have an appreciable dependence on $v$ (and thus on $L$) for very large values of $v$, the fit indicates that $f_\infty=\lim_{v\to\infty}f(v)$ is different from zero. 
The attempts to fit $f(v)$ with functions vanishing for large $v$ yield poor results.
In addition we have  explicitly verified that the inclusion of the quartic terms in the free-energy gives a subleading contribution to $\si$. 

\section{Conclusions} 
In this paper we have considered the interface free-energy obtained by the imposition of  different boundary conditions on two identical copies of a spin glass. We have concentrated our study to the behavior of the free energy difference with the size of the system at $T_c$. We have found that the behavior of the interface follows the usual pattern of critical phenomena.  The free-energy difference above 6 dimensions is of order $L^{D-6}$, or $\sigma \sim L^{-5}$ as suggested by simple  scaling laws.  This means that RSB effects may be neglected to  leading order at the critical temperature. This is very different from what found at lower temperatures \cite{FPVint}, where the scaling of the interface was found to depend critically on the zero modes associated with RSB.  
We expect that below dimension 6 the free-energy difference remains finite (or $\sigma \sim L^{D-1}$), a prediction that can be tested in numerical simulations. Indeed since there does not seem to be any RSB at $T_c$, the usual arguments namely a) the singular part of the free energy scales like a correlation volume, b) finite size scaling, should presumably apply.

\newpage
\appendix {\bf{ Appendix A  : Pure systems}}

The free energy is
\be F = A \int_{-L/2} ^{+L/2} dz\  [ \frac {1}{2} \vf'^2 + \frac{u}{4} \vf^4] \ee
\begin{enumerate}
\item {Up-up boundary conditions}

The magnetization m in the planes $z=\pm L/2$ is given  and we solve the equation of motion
\be - \vf" + u \vf^3=0 \ee
with the b.c.
\be \vf (-L/2) = \vf (+L/2) = m \ee
The mechanical analog is a negative energy bounce off the potential $-\frac{u}{4}\vf^4$ starting at $\vf =m $ bouncing at $\vf_0$ and returning to $\vf =m$. It is given  by
 \be \int_m^{\vf(z)} \frac{d\psi}{ \sqrt{ \psi^4- \vf_0^4}} = -\sqrt{\frac{u}{2}} (L/2+z) \ee
 Then $\vf_0$ is determined by
 \be   \int_m^{\vf_0} \frac{d\psi}{ \sqrt{ \psi^4- \vf_0^4}} = -\sqrt{\frac{u}{8}} L \ee
For  $L$ large  the order parameter $\vf_0$ at the centre of the sample is small, much smaller than the given finite value $m$ on the boundaries,  and it is thus given asymptotically by 
\be L \vf_0 = \sqrt{\frac{8}{u}}  \int _1^{\infty} \frac {dt}{ \sqrt{t^4-1} } + O(1/L) \ee
The free energy is then given by 
\ba\frac{ F_{\uparrow \uparrow}}{A} && = \frac{u}{2} \int _{-L/2} ^{+L/2} dz (\vf^4-\vf_0^4) + \frac{u}{4} L \vf_0^4 \nonumber \\&&= - \sqrt{2u} \int_m^{\vf_0} d\psi \sqrt{ \psi^4-\vf_0^4}  + \frac{u}{4} L \vf_0^4 \nonumber \\
&&= \frac {1}{3} \sqrt{2u} m^3 -\sqrt{2u} \frac{\vf_0^3}{3} + \frac{u}{4} L\vf_0^4 + \sqrt {2u}\vf_0 ^3 \int_1^{\infty} \frac {dt} { \sqrt{t^4-1} +t^2}.\ea
The corrections to the first leading term are of order $1/L^3$. We finally note that
\be \int_1^{\infty} \frac {dt} {\sqrt{t^4-1}} = \frac {\Gamma^2(1/4)}{ \sqrt{32 \pi} }\ee
and
\be  \int_1^{\infty} \frac {dt} { \sqrt{t^4-1} +t^2} = \frac {\Gamma^2(1/4)}{ \sqrt{72 \pi}} -\frac{1}{3}.
 \ee
 \item {Up-down boundary conditions}
 
 We are still dealing with the mechanical analog of a motion in the inverted potential $-u\vf^4/4$ but, now with a positive energy solution going from  $\vf(L/2) =m$ to  $\vf(-L/2) =-m$. The solution is  given by
 \be \int _m^\vf \frac {d\psi}{ \sqrt{ \psi^4+ \vf_1^4}} = -\sqrt{\frac{u}{2}} (\frac {L}{2} -z) \ee
 and $\vf_1$ is determined by
 \be \int _{-m}^{+m} \frac {d\psi}{ \sqrt{ \psi^4+ \vf_1^4}} = \sqrt{\frac{u}{2}} L\ee
Again $\vf_1$ is of order $1/L$ and given asymptotically by
\be L\vf_1 = \sqrt{\frac{2}{u}} \int_{-\infty}^{+\infty} \frac{dt}{\sqrt{ t^4+1}} \ee

Then the free energy 
\be\frac{ F_{\uparrow, \downarrow}}{A} = -\frac{u}{4} L \vf_1^4 +  \sqrt{2u}\vf_1^3  \int_{0}^{+m/\vf_1}dt \sqrt{t^4+1} \ee
The last asymptotic estimate 
\be \int_{0}^{+m/\vf_1}dt \sqrt{t^4+1} = \frac{1}{3}( \frac{m}{\vf_1})^3 + \int_0^{\infty} \frac {dt}{\sqrt{t^4+1} +t^2} \ee
and the values of the integrals
\be \int_{-\infty}^{+\infty} \frac{dt}{\sqrt{ t^4+1}} =  \frac {\Gamma^2(1/4)}{ \sqrt{16 \pi}} \ee
\be \int_0^{\infty} \frac {dt}{\sqrt{t^4+1} +t^2} = \frac {\Gamma^2(1/4)}{ \sqrt{36 \pi}} \ee
completes the calculation. 

\end{enumerate}

We may now compute $ F_{\uparrow, \downarrow} - F_{\uparrow, \uparrow}$ ; the constant $m^3$ term cancels and we are left with a difference proportional to $A/L^3$ as expected 
\be \frac {  F_{\uparrow, \downarrow} - F_{\uparrow, \uparrow}}{A} = \frac{1}{u L^3}  [ \frac{5}{3\cdot 2^{10} \pi^2 }\Gamma^8(1/4) - \frac {1}{3\cdot 4^3 \pi \sqrt{2\pi}} \Gamma^6(1/4) ] \ee
\vskip 2mm
\appendix {\bf{ Appendix B  : A spherical Spin Glass Model }}

\begin{figure}[ht]
\begin{center}
\includegraphics[width= 0.7 \textwidth]{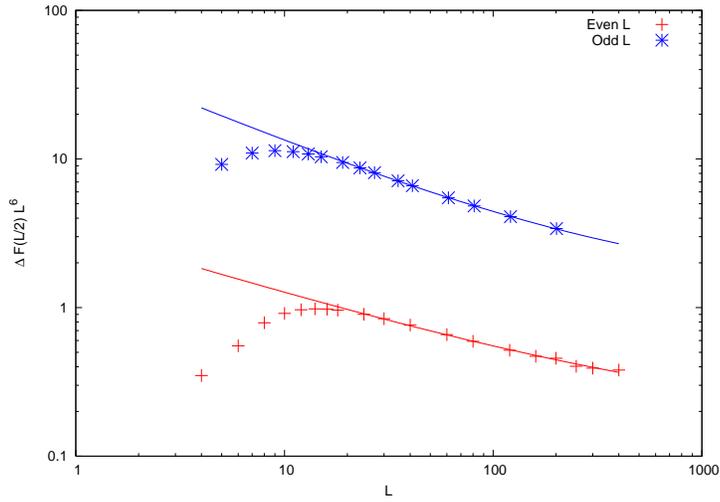}
\caption{The 
rescaled difference of free-energy density in the center of the box for the two kinds of boundary conditions $\si (L/2) L^6$ as a function of the system size $L$. Notice that even and odd values of $L$ give rise to different curves.  The points are plotted together with a power law fit of the form $f(L)=a+\frac {b}{L^c}$ (solid lines). Chi-square fitting for $L>20$ gives the parameters $a=0.15\pm 0.04$, $b=3.1\pm 0.3$ and $c=0.44\pm 0.04$ for $L$ even and $a=1.3\pm 0.1$, $b=46.9\pm 1.7$ and $c=0.58\pm 0.04$ for $L$ odd. }
\label{sph}
\end{center}
\end{figure}

As  observed in the main text, the boundary conditions defined in ref. \cite{CGGPV}, imply high overlaps on the boundaries, and they cannot be analyzed in the context of Landau small order parameter expansions of the free-energy. We define here a model with the following properties: 
\begin{itemize}
\item The replica free-energy of the model admits a simple closed form in terms of the overlap matrix $Q_{a,b}(x)$, and it can be explicitly  continued to the zero replica limit $n\to 0$,  when the Parisi ansatz is assumed.  
\item The Landau expansion close to $T_c$ coincides, up to the fourth order term,
with that of the SK model. 
\end{itemize}
Consider a system where on each site, labelled by its longitudinal and transverse coordinates  $x$ and ${\bf y}$ respectively,  
 there are $K$ spins $\sigma_{x,{\bf y}}^r$, $r=1,...,K$,  subject to the spherical constraint 
$\sum_r  (\sigma_{x,{\bf y}}^r)^2=K$. 
We write the Hamiltonian of the model as a sum of terms that couple spins on the same 
plane and terms that couple spins in adjacent planes. 
\begin{eqnarray}
H=\sum_{x=0}^L {\cal H}_{x}^{ort}
+
\sum_{x=0}^{L-1} 
 {\cal H}_{x}^{par}
\label{hamiltonian}
\end{eqnarray}
where the Hamiltonians  ${\cal H}_{x}^{ort}$ and 
$ {\cal H}_{x}^{par}$ are gaussian random variables with variances
\begin{eqnarray}
\langle {\cal H}_{x}^{ort}[\sigma]{\cal H}_{x}^{ort}[\tau]\rangle
&=&
\sum_{\bf y} f_x(
\frac{1}{2^{d-1}}
\sum_{\bf z\in V_{\bf y}} q_{{\bf y},{\bf z}}^{x,x}) \nonumber\\
\langle {\cal H}_{x}^{par}[\sigma]{\cal H}_{x}^{par}[\tau]\rangle&=&\sum_{\bf y}
f(q_{{\bf y},{\bf y}}^{x,x+1}) \nonumber\\
\end{eqnarray}
where we denoted by $\sigma^x_{\bf y}$ the value of the spin on site ${\bf
  y}$ of the $x$-th plane, $q_{{\bf y},{\bf z}}^{x,w}=\frac 1 K
\sum_{r=1}^K \sigma_{x,{\bf y}}^r\tau_{w,{\bf z}}^r$, the overlaps
between spin configurations $\sigma$ and $\tau$ on different sites. The different  functions
of the overlap are chose to be 
$f(q)=\frac 1 2 (q^2+ y q^4)$, $f_x( q)=f(q)$ for $x\ne 0,L$ and
$f(0)=\frac 1 4 f(q)$ for $x=0,L$. Notice that with this choice,  it is possible
to express  the Hamiltonian (\ref{hamiltonian})
in terms of two body and four body Gaussian couplings
between the spins. We can now introduce two copies of the system with up-up and up-down boundary conditions. The up-up conditions consist
in considering two copies $\sigma_{x,{\bf y}}^r$ and $\tau_{x,{\bf
y}}^r$ constrained to be identical for $x=0,L$: $\sigma_{0,{\bf
y}}^r=\tau_{0,{\bf y}}^r$ and $\sigma_{L,{\bf y}}^r=\tau_{L,{\bf
y}}^r$. The up-down conditions consist
of  two copies $\sigma_{x,{\bf y}}^r$ and $\tau_{x,{\bf
y}}^r$ constrained to be identical for $x=0$ but with opposite values for $x=L$: $\sigma_{0,{\bf
y}}^r=\tau_{0,{\bf y}}^r$ and $\sigma_{L,{\bf y}}^r=-\tau_{L,{\bf
y}}^r$. As for the reduced model, the replica treatment of the problem involves the introduction of two local $n\times n$ overlap matrices, which under the replica symmetric ansatz and the assumption of independence of the overlap profiles of the transverse spatial coordinate, can be parametrized in terms of the functions $q(x)$, $l(x)$ and $k(x)$ $x=0,...,L$ of the main text.  
The resulting $F_{\uparrow, \uparrow}$ free-energies and $F_{\uparrow, \downarrow}$ as a function of these parameters can be decomposed as
 \begin{eqnarray}
&& F_{\uparrow, \uparrow}[\{q,l,k\}]= F^{bulk}[\{q,l,k\}]+F^{boundary}_0+F^{boundary}_{\uparrow, \uparrow}\\
&& F_{\uparrow, \downarrow}[\{q,l,k\}]= F^{bulk}[\{q,l,k\}]+F^{boundary}_0+F^{boundary}_{\uparrow, \downarrow}
\end{eqnarray}
where
\begin{eqnarray}
&&-\beta F^{bulk}=\beta^2\sum_{x=1}^{L-1}[ f(1) +f(l(x))-f(q(x))-f(k(x))]\nonumber \\
&& +
\beta^2\sum_{x=1}^{L-2} f(1) +f((l(x)+l(x+1))/2)-f((q(x)+q(x+1))/2)-f((k(x)+k(x+1))/2)\nonumber \\
&&+\frac 1 2 \sum_{x=1}^{L-1}
\Log\left[
1+l(x)-q(x)-k(x)
\right]
+\frac{q(x)+k(x)}{1+l(x)-q(x)-k(x)}
\nonumber \\&&
+\Log\left[
1-l(x)-q(x)+k(x)
\right]
+\frac{q(x)-k(x)}{1-l(x)-q(x)+k(x)}
\end{eqnarray}

\begin{eqnarray}
&&-\beta F^{boundary}_{0}=
\frac 1 2 \beta^2( 2f(1) -f(q(0))-f(q(L))\nonumber \\
&&+\beta^2(f(1) +f((1+l(1))/2)-f((q(1)+q(0))/2)-f((q(0)+k(1))/2)\nonumber \\
&&
+ \frac 1 2\left[
\Log\left[
1-q(0)
\right]
+\frac{q(0)}{1-q(0)}
+ 
\Log\left[
1-q(L)
\right]
+\frac{q(L)}{1-q(L)}
\right] 
\end{eqnarray}

\begin{eqnarray}
&&-\beta F^{boundary}_{\uparrow, \uparrow}=\nonumber\\
&&+\beta^2(f(1) +f((1+l(L-1))/2)-f((q(L-1)+q(L))/2)-f((q(L)+k(L-1))/2)\nonumber \\
\end{eqnarray}
\begin{eqnarray}
&&-\beta F^{boundary}_{\uparrow, \downarrow}=\nonumber\\
&&+\beta^2(f(1) +f((1+l(L-1))/2)-f((q(L-1)+q(L))/2)-f((q(L)-k(L-1))/2)\nonumber \\
\end{eqnarray}
Given the complexity of the expression, we have manipulated them through the 
Mathematica software in order to obtain the equations of motion, and integrated the resulting equations numerically by the relaxation method. 
As a proxy of the free-energy difference 
$F_{\uparrow, \downarrow}-F_{\uparrow, \uparrow}$,  in figure \ref{sph} we plot
 $\si (L/2) L^6$, 
the  difference in free-energy density in the center of the box multiplied by $L^6$, which according to the argument given in the main text should tend to a constant for large $L$.
 The numerical error on $\si (L/2)$, which is an extremely small quantity, limited the range of $L$ that we could investigate to $L\leq 400$.  
The integration gives result compatible with the analysis of the reduced model, confirming the independence on the 
detailed boundary conditions imposed. 
As in the case of the reduced model, 
the curves can be fitted by the form
 $f(L)=a+b/L^c$, though in this case a logarithmic fit of the form $f(L)=a/\log(L)^c$
give a fit of comparable quality. 

\end{document}